\documentclass[reqno]{amsart}

\usepackage{IEEEtrantools}
\usepackage{blindtext}
\usepackage{tasks}
\usepackage{amssymb}
\usepackage{amsmath,latexsym}
\usepackage{graphicx}
\usepackage{mathrsfs}
\usepackage{dsfont}
\usepackage{amsfonts}
\usepackage{breqn}
\usepackage{mathtools}
\usepackage{enumitem}   
\usepackage{mathrsfs}
\usepackage{url} 

\usepackage{algorithm}
\usepackage{algorithmic}

\usepackage{tikz}
\usetikzlibrary{calc,shapes,arrows}

\usepackage{xspace}
\newcommand{\software}{\textsf{FAUST}$^{\mathsf 2}$\xspace}

\newtheorem{theorem}{Theorem}[section]

\newtheorem{definition}[theorem]{Definition}

\newtheorem{remark}[theorem]{Remark}
\newtheorem{assumption}[theorem]{Assumption}
\numberwithin{equation}{section}

\newcommand{\intcc}[1]{\ensuremath{{\left[#1\right]}}}

\newcommand{\R}{{\mathbb{R}}}

\newcommand{\N}{{\mathbb{N}}}

\newcommand{\Let}{:=}
\newcommand{\EE}{\mathds{E}}

\begin{document}

\begin{abstract}
This paper provides a compositional scheme based on dissipativity approaches for constructing finite abstractions of continuous-time continuous-space stochastic control systems. The proposed framework enjoys the structure of the interconnection topology and employs a notion of \emph{stochastic storage functions}, that describe joint dissipativity-type properties of subsystems and their abstractions. By utilizing those stochastic storage functions, one can establish a relation between continuous-time continuous-space stochastic systems and their finite counterparts while quantifying probabilistic distances between their output trajectories. Consequently, one can employ the finite system as a suitable substitution of the continuous-time one in the controller design process with a guaranteed error bound. In this respect, we first leverage dissipativity-type compositional conditions for the compositional quantification of the distance between the interconnection of continuous-time continuous-space stochastic systems and that of their discrete-time (finite or infinite) abstractions. We then consider a specific class of stochastic affine systems and construct their finite abstractions together with their corresponding stochastic storage functions. The effectiveness of the proposed results is demonstrated by applying them to a temperature regulation in a circular network containing $100$ rooms and compositionally constructing a discrete-time abstraction from its original continuous-time dynamic. The constructed  discrete-time abstraction is then utilized as a substitute to compositionally synthesize policies keeping the temperature of each room in a comfort zone.
\end{abstract}

\title[Compositional Construction of Finite MDPs for Continuous-Time Stochastic Systems]{Compositional Construction of Finite MDPs for Continuous-Time Stochastic Systems: A Dissipativity Approach}

\author{Ameneh Nejati$^1$}
\author{Majid Zamani$^{2,3}$}
\address{$^1$Department of Electrical and Computer Engineering, Technical University of Munich, Germany.}
\email{amy.nejati@tum.de}
\address{$^2$Department of Computer Science, University of Colorado Boulder, USA}
\address{$^3$Department of Computer Science, Ludwig Maximilian University of Munich, Germany}
\email{majid.zamani@colorado.edu}
\maketitle

\section{Introduction}
{\bf Motivations.} Automated controller synthesis for continuous-time continuous-space stochastic systems against high-level logical properties such as those expressed as linear temporal logic (LTL) formulae~\cite{pnueli1977temporal} is naturally a difficult task mainly due to continuous state sets. To deal with this problem, one potential direction is to first abstract the given system by a simpler one, \emph{i.e.,} discrete in time and potentially in space, then synthesize a desired controller for the abstract system, and finally transfer the controller back to the original one while quantifying probabilistic error bounds.

Unfortunately, \emph{curse of dimensionality} is the main problem in the construction of finite abstractions (a.k.a. finite Markov decision processes (MDPs)) for large-scale systems: the complexity of constructing finite abstractions  increases exponentially with the dimension of the state set. Compositional techniques play significant roles to alleviate this complexity. In this regard, one can consider the large-scale stochastic system as an interconnected system composed of several smaller subsystems, and then develop a compositional scheme for the construction of finite abstractions for the given complex system via abstractions of smaller subsystems.

\smallskip

{\bf Related Literature.} There have been some results, proposed in the past few years, on the construction of finite abstractions for \emph{continuous-time} continuous-space stochastic systems. A reachability analysis for continuous-time stochastic systems by constructing Markov chain with quantified error bounds is proposed in~\cite{laurenti2017reachability}. Abstraction approaches for incrementally stable stochastic control systems without discrete dynamics, incrementally stable stochastic switched systems, and randomly  switched stochastic systems are respectively studied in~\cite{zamani2014symbolic},~\cite{zamani2015symbolic}, and~\cite{zamani2014approximately}. Although original systems in~\cite{zamani2014symbolic},~\cite{zamani2015symbolic}, and~\cite{zamani2014approximately} are stochastic, their abstractions are constructed as finite labeled transition systems while finite abstractions in this work are presented as finite Markov decision processes. Finite labeled transition systems in this context are useful only if the noise in the system is small. An approximation scheme for the construction of infinite abstractions for jump-diffusion processes is developed in~\cite{julius2009approximations}. Compositional construction of infinite abstractions via small-gain type conditions is proposed in~\cite{zamani2016approximations}. An (in)finite abstraction-based technique for synthesis of continuous-time stochastic control systems is recently discussed in~\cite{Nejati2019ECC}.

For \emph{discrete-time} stochastic systems with continuous-state sets, there also exist several results. Finite abstractions for formal synthesis of discrete-time stochastic control systems are proposed in~\cite{APLS08}. An adaptive and sequential gridding approach is proposed in~\cite{SSoudjani}, and~\cite{SA13} with dedicated tools \software~\cite{FAUST15} and \emph{StocHy}~\cite{StocHy}. Moreover, formal abstraction-based policy synthesis is discussed in~\cite{tmka2013}, and~\cite{Kamgarpour13}. Compositional construction of infinite abstractions via classic small-gain and dissipativity conditions is respectively proposed in \cite{lavaei2017compositional,lavaei2018CDCJ}. Compositional construction of finite abstractions utilizing dynamic Bayesian networks and dissipativity conditions is studied in~\cite{SoudjaniAR2017} and~\cite{lavaei2017HSCC}, respectively. Although the proposed compositional approach in~\cite{lavaei2017HSCC} is based on dissipativity conditions, their results are provided for discrete-time systems. In comparison, we deal with continuous-time systems here and the ultimate goal is to develop a compositional approach to construct finite MDPs from \emph{continuous-time} stochastic systems.

Compositional construction of (in)finite abstractions via $\max$-type small-gain conditions is proposed in \cite{lavaei2018ADHS,lavaei2018ADHSJ}. Compositional construction of finite abstractions for networks of stochastic systems via \emph{relaxed} small-gain and dissipativity approaches is respectively presented in~\cite{lavaei2019ECC,lavaei2019NAHS}. A notion of approximate simulation relation for stochastic systems based on a lifting probabilistic evolution of systems is proposed in \cite{SSA16}. This notion is generalized in~\cite{lavaei2019NAHS1} for compositional abstraction-based synthesis of general MDPs. Compositional construction of finite abstractions for networks of stochastic \emph{switched} systems accepting multiple Lyapunov functions with dwell-time conditions is presented in~\cite{lavaei2019HSCC_J,lavaei2019LSS} via respectively small-gain and dissipativity approaches. Compositional construction of infinite and finite abstractions for large-scale \emph{discrete-time} stochastic systems via different novel compositionality conditions is widely discussed in~\cite{lavaei2019Thesis}.

{\bf Contributions.} In this paper, we provide a compositional scheme for constructing finite MDPs from continuous-time continuous-space stochastic systems.
We derive dissipativity-type conditions to propose compositionality results which are established based on relations between continuous-time subsystems and that of their abstract counterparts utilizing notions of so-called \emph{stochastic storage functions}. The provided compositionality conditions can enjoy the structure
of interconnection topology and be potentially fulfilled independently
of the interconnection or gains of the subsystems (cf. the case study).

To this end, we first compositionally quantify the probabilistic distance between the interconnection of continuous-time continuous-space stochastic subsystems and their discrete-time (finite or infinite) abstractions. We then focus on a particular class of stochastic affine systems and construct their finite abstractions together with their corresponding stochastic storage functions. Finally, we illustrate the effectiveness of the proposed techniques by applying them to a physical case study.

{\bf Recent Works.}
Compositional abstraction-based synthesis of continuous-time stochastic systems is also proposed in~\cite{Nejati2019ECC_J}, but using a different compositionality scheme based on \emph{small-gain} conditions. Our proposed compositionality approach here can be potentially less conservative than the one presented in~\cite{Nejati2019ECC_J} for some classes of systems. The dissipativity-type compositional reasoning proposed here can enjoy the structure of the interconnection topology and may not require any constraint on the number or gains of subsystems (cf. Remark~\ref{compositionality remark} and the case study). Consequently, the proposed approach here can provide a scale-free compositionality condition which is independent of the number of subsystems compared to the proposed results in~\cite{Nejati2019ECC_J}.

\section{Notations and Model Classes}\label{Preliminaries}

\subsection{Notations}
A probability space in this work is defined as $(\Omega,\mathcal F_{\Omega},\mathds{P}_{\Omega})$,
where $\Omega$ is the sample space,
$\mathcal F_{\Omega}$ is a sigma-algebra on $\Omega$ comprising subsets of $\Omega$ as events, and $\mathds{P}_{\Omega}$ is a probability measure that assigns probabilities to events. We assume that triple $(\Omega,\mathcal F_{\Omega},\mathds{P}_{\Omega})$ denotes a probability space endowed with a filtration $\mathbb{F} = (\mathcal F_s)_{s\geq 0}$ satisfying the usual conditions of completeness and right continuity.

Sets of nonnegative and positive integers are respectively denoted by $\mathbb N := \{0,1,2,\ldots\}$ and $\mathbb N_{\ge 1} := \{1,2,3,\ldots\}$. Symbols $\mathbb R$, $\mathbb R_{>0}$, and $\mathbb R_{\ge 0}$ respectively denote sets of real, positive and nonnegative real numbers. We use $x = [x_1;\ldots;x_N]$ to denote the corresponding vector of dimension $\sum_i n_i$, given $N$ vectors $x_i \in \mathbb R^{n_i}$, $n_i\in \mathbb N_{\ge 1}$, and $i\in\{1,\ldots,N\}$.
Given functions $f_i:X_i\rightarrow Y_i$, for any $i\in\{1,\ldots,N\}$, their Cartesian product $\prod_{i=1}^{N}f_i:\prod_{i=1}^{N}X_i\rightarrow\prod_{i=1}^{N}Y_i$ is defined as $(\prod_{i=1}^{N}f_i)(x_1,\ldots,x_N)=[f_1(x_1);\ldots;f_N(x_N)]$.
We denote by $\Vert\cdot\Vert$ the Euclidean norm. Given a function $f:\mathbb N\rightarrow\mathbb{R}^n$, the supremum of $f$ is denoted by $\Vert f\Vert_{\infty} \Let \text{(ess)sup}\{\Vert f(k)\Vert,k\geq 0\}$.
The identity matrix in $\mathbb R^{n\times{n}}$  is denoted by $\mathds{I}_n$. Column vectors in $\mathbb R^{n\times{1}}$ with all elements equal to zero and one are respectively denoted by $\mathbf{0}_n$ and $\mathds{1}_n$. A function $\gamma:\mathbb\mathbb \mathbb R_{\ge 0}\rightarrow\mathbb\mathbb \mathbb R_{\ge 0}$, is said to be a class $\mathcal{K}$ function if it is continuous, strictly increasing, and $\gamma(0)=0$.
A class $\mathcal{K}$ function $\gamma(\cdot)$ is said to be a class $\mathcal{K}_{\infty}$ if
$\gamma(s) \rightarrow \infty$ as $s\rightarrow\infty$.

\subsection{Continuous-Time Stochastic Control Systems}
\begin{definition}
	A continuous-time stochastic control system (ct-SCS) in this paper is defined by the tuple
	\begin{equation}\label{eq:ct-SCS}
	\Sigma=(X,U,W,\mathcal U,\mathcal W,f,\sigma,Y_1, Y_2,h_1,h_2),
	\end{equation}
	where: 
	\begin{itemize}
		\item $X\subseteq \mathbb R^n$ is the state set of the system;
		\item $U\subseteq \mathbb R^m$ is the \emph{external} input set of the system;
		\item $W\subseteq \mathbb R^p$ is the \emph{internal} input set of the system;
		\item $\mathcal U$ and $\mathcal W$ are subsets of the sets of all $\mathbb{F}$-progressively measurable processes taking values respectively in $\mathbb R^m$ and $\mathbb R^p$; 
		\item $f:X\times U \times W\rightarrow X$ is the drift term which is globally Lipschitz continuous: there exist constants $\mathscr{L}_x, \mathscr{L}_\nu, \mathscr{L}_w \in  \mathbb R_{\ge0}$ such that $\Vert f(x,\nu,w)-f(x',\nu',w')\Vert \leq \mathscr{L}_x\Vert x-x'\Vert+\mathscr{L}_\nu\Vert \nu-\nu'\Vert+\mathscr{L}_w\Vert w-w'\Vert$ for all $x,x' \in X$, for all $\nu,\nu' \in U$, and for all $w,w' \in W$;
		\item $\sigma: \mathbb R^n \rightarrow \mathbb R^{n\times \textsf b}$ is the diffusion term which is globally Lipschitz continuous with the Lipschitz constant $\mathscr{L}_\sigma$;
		\item  $Y_1\subseteq \mathbb R^{q_1}$ is the \emph{external} output set of the system;
		\item  $Y_2\subseteq \mathbb R^{q_2}$ is the \emph{internal} output set of the system;
		\item  $h_1:X\rightarrow Y_1$ is the \emph{external} output map;
		\item  $h_2:X\rightarrow Y_2$ is the \emph{internal} output map.
	\end{itemize}
\end{definition}	
A continuous-time stochastic control system $\Sigma$ satisfies
\begin{align}\label{Eq_1}
\Sigma:\left\{\hspace{-1mm}\begin{array}{l}\mathsf{d}\xi(t)=f(\xi(t),\nu(t),w(t))\,\mathsf{d}t+\sigma(\xi(t))\,\mathsf{d}\mathbb W_t,\\
\zeta_1(t)=h_1(\xi(t)),\\
\zeta_2(t)=h_2(\xi(t)),\\
\end{array}\right.
\end{align}    
$\mathds P$-almost surely ($\mathds P$-a.s.) for any $\nu \in \mathcal U$ and $w \in \mathcal W$, where $(\mathbb W_t)_{t \ge 0}$ is a ${\textsf b}$-dimensional Brownian motion, and stochastic processes $\xi:\Omega \times \mathbb R_{\ge 0}\rightarrow X$, $\zeta_1:\Omega \times \mathbb R_{\ge 0}\rightarrow Y_1$, and $\zeta_2:\Omega \times \mathbb R_{\ge 0}\rightarrow Y_2$ are respectively called the \emph{solution process} and the external and internal \emph{output trajectories} of $\Sigma$. We also use $\xi_{a \nu w}(t)$ to denote the value of the solution process at time $t\in\mathbb R_{\ge 0}$ under input trajectories $\nu$ and $w$ from an initial condition $\xi_{a \nu w}(0)= a$ $\mathds P$-a.s., where $a$ is a random variable that is $\mathcal F_0$-measurable. We also denote by $\zeta_{1_{a \nu w}}$ and $\zeta_{2_{a \nu w}}$ the external and internal \emph{output trajectories} corresponding to the \emph{solution process} $\xi_{a \nu w}$.

\begin{remark}
Note that in this article, the term ``internal'' is used for inputs and outputs of subsystems that are affecting each other in the interconnection topology while properties of interest are defined over ``external'' outputs. The ultimate goal is to synthesize ``external'' inputs to fulfill desired properties over ``external'' outputs.
\end{remark}
In this paper, we are interested in investigating interconnected continuous-time stochastic systems, defined later in Subsection~\ref{Interconnected SCS}, without internal signals. Then the tuple~\eqref{eq:ct-SCS} reduces to $(X,U,\mathcal U,f,\sigma,Y, h)$ with $f:X\times U\rightarrow X$, and ct-SCS~\eqref{Eq_1} can be re-written as
\begin{align}\notag
\Sigma:\left\{\hspace{-1mm}\begin{array}{l}\mathsf{d}\xi(t)=f(\xi(t),\nu(t))\,\mathsf{d}t+\sigma(\xi(t))\,\mathsf{d}\mathbb W_t,\\
\zeta(t)=h(\xi(t)).\\
\end{array}\right.
\end{align}

\subsection{Finite Abstractions of ct-SCS}\label{subsec:MDP}

In order to construct finite abstractions of continuous-time stochastic systems, we first need to provide a \emph{time-discretized} version of ct-SCS in~\eqref{Eq_1} as in the following definition.

\begin{definition}	
	A \emph{time-discretized} version of ct-SCS $\Sigma$ is defined by the tuple
	\begin{equation}\label{eq:dt-SCS}
	\widetilde \Sigma=\left(\tilde X,\tilde U, \tilde W, \varsigma,\tilde f,\tilde Y_1,\tilde Y_2,\tilde h_1,\tilde h_2\right)\!,
	\end{equation}
	where: 
	\begin{itemize}
		\item $\tilde X\subseteq \mathbb R^n$ is a Borel space as the state set of the system. We denote by $(\tilde X, \mathcal B (\tilde X))$ the measurable space with $\mathcal B(\tilde X)$  being  the Borel sigma-algebra on the state space;
		\item $\tilde U\subseteq \mathbb R^m$ is a Borel space as the \emph{external} input set;
		\item $\tilde W\subseteq \mathbb R^p$ is a Borel space as the \emph{internal} input set;
		\item $\varsigma$ is a sequence of independent and identically distributed (i.i.d.) random variables from a sample space $\Omega$ to the set $\mathcal V_\varsigma$,
		\begin{equation*}
		\varsigma:=\{\varsigma(k):\Omega\rightarrow \mathcal V_{\varsigma},\,\,k\in\mathbb N\};
		\end{equation*}
		\item  $\tilde f:\tilde X\times \tilde U\times\tilde W \times \mathcal V_{\varsigma} \rightarrow \tilde X$ is a measurable function characterizing the state evolution of the system;
		\item $\tilde Y_1\subseteq \mathbb R^{q_1}$ is a Borel space as the \emph{external} output set;
		\item $\tilde Y_2\subseteq \mathbb R^{q_2}$ is a Borel space as the \emph{internal} output set;
		\item $\tilde h_1:\tilde X\rightarrow \tilde Y_1$ is the \emph{external} output map;
		\item $\tilde h_2:\tilde X\rightarrow \tilde Y_2$ is the \emph{internal} output map.
	\end{itemize}
\end{definition}
The evolution of $\widetilde\Sigma$, for given initial state $\tilde x(0)\in \tilde X$ and input sequences $\{\tilde\nu(k):\Omega\rightarrow \tilde U,\,\,k\in\mathbb N\}$ and $\{\tilde w(k):\Omega\rightarrow \tilde W,\,\,k\in\mathbb N\}$, can be written as
\begin{align}\label{discrete-version}
\widetilde\Sigma:\left\{\hspace{-1mm}\begin{array}{l}\tilde \xi(k+1)=\tilde f(\tilde \xi(k),\tilde \nu(k),\tilde w(k),\varsigma(k)),\\
\tilde \zeta_1(k)=\tilde h_1(\tilde \xi(k)),\\
\tilde \zeta_2(k)=\tilde h_2(\tilde \xi(k)),\\
\end{array}\right.
k\in\mathbb N.
\end{align}
The sets $\mathcal {\tilde U}$ and $\mathcal {\tilde W}$ are associated to $\tilde U$ and $\tilde W$ to be the collections of sequences $\{\tilde\nu(k):\Omega\rightarrow \tilde U,\,\,k\in\mathbb N\}$ and $\{\tilde w(k):\Omega\rightarrow \tilde W,\,\,k\in\mathbb N\}$, in which $\tilde\nu(k)$ and $\tilde w(k)$ are independent of $\varsigma(z)$ for any $k,z\in\mathbb N$ and $z\ge k$. For any initial state $\tilde a\in \tilde X$, $\tilde \nu(\cdot)\in\mathcal{\tilde U}$ and $\tilde w(\cdot)\in\mathcal{\tilde W}$, the random sequences $\tilde \xi_{\tilde a\tilde \nu \tilde w}:\Omega \times\mathbb N \rightarrow \tilde X$, $\tilde \zeta_{1_{\tilde a\tilde \nu \tilde w}}:\Omega \times \mathbb N \rightarrow \tilde Y_1$, and $\tilde \zeta_{2_{\tilde a\tilde \nu \tilde w}}:\Omega \times \mathbb N \rightarrow \tilde Y_2$ fulfilling~\eqref{discrete-version} are respectively called the \textit{solution process}, and external and internal \textit{output trajectories} of $\widetilde\Sigma$ under an external input $\tilde \nu$, an internal input $\tilde w$, and an initial state $\tilde a$. 

\begin{remark}
	Note that the discrete-time system $\tilde\Sigma$ in~\eqref{discrete-version} is presented independently of ct-SCS $\Sigma$ for now. In particular, in order to construct finite abstractions of continuous-time stochastic systems $\Sigma$ (i.e., $\widehat\Sigma$) as proposed in Definition~\ref{finite system}, one first needs to provide a time-discretized version of ct-SCS (i.e., $\widetilde\Sigma$) as a middle stage. In Section~\ref{sec:constrcution_finite}, we focus on a particular class of continuous-time stochastic affine systems $\Sigma$ and discuss the best choice for $\widetilde\Sigma$ to acquire the least approximation error between $\Sigma$ and $\widehat\Sigma$.
\end{remark}

The discrete-time stochastic control system $\widetilde\Sigma$ can be \emph{equivalently} reformulated as a Markov decision process~\cite[Proposition 7.6]{kallenberg1997foundations}
\begin{equation}\notag
\widetilde \Sigma=\left(\tilde X,\tilde U, \tilde W, \tilde T_{\mathsf {\tilde x}},\tilde Y_1,\tilde Y_2,\tilde h_1,\tilde h_2\right)\!,	
\end{equation}
where the map $\tilde T_{\mathsf {\tilde x}}:\mathcal B(\tilde X)\times \tilde X\times \tilde U\times \tilde W\rightarrow[0,1]$,
is a conditional stochastic kernel that assigns to any $\tilde x \in \tilde X$, $\tilde \nu\in \tilde U$, and $\tilde w\in \tilde W$, a probability measure $\tilde T_{\mathsf {\tilde x}}(\cdot | \tilde x,\tilde \nu, \tilde w)$
on the measurable space
$(\tilde X,\mathcal B(\tilde X))$
so that for any set $\mathcal A \in \mathcal B(\tilde X)$, 
\begin{align}\notag
\mathds P (\tilde x(k+1)\in \mathcal A\,\big|\,\tilde x(k),\tilde \nu(k),\tilde w(k)) =& \int_{\mathcal A} \tilde T_{\mathsf {\tilde x}} (\mathsf{d}\tilde x(k+1)\,\big|\,\tilde x(k),\tilde \nu(k),\tilde w(k)).
\end{align}
For given inputs $\tilde\nu(\cdot), \tilde w(\cdot),$ the stochastic kernel $\tilde T_{\mathsf {\tilde x}}$ captures the evolution of the state of $\widetilde\Sigma$  and can be uniquely specified by the pair $(\varsigma,\tilde f)$ from \eqref{eq:dt-SCS}. We now define \emph{Markov policies} in order to control the system.

\begin{definition}
	For the discrete-time stochastic control system $\widetilde\Sigma$ in~\eqref{discrete-version}, a Markov policy  is a sequence $\bar\mu = (\bar\mu_0,\bar\mu_1,\bar\mu_2,\ldots)$ of universally measurable stochastic kernels $\bar\mu_n$ \cite{BS96}, each defined on the input space $\tilde U$ given $\tilde X \times \tilde W$ such that for all $(\tilde \xi_n, \tilde w_n)\in \tilde X \times \tilde W$, $\bar\mu_n(\tilde U|(\tilde \xi_n, \tilde w_n))=1$. The class of all such Markov policies is denoted by $\mathcal P_{\mathcal M}$. 
\end{definition}

Now we construct finite MDPs $\widehat\Sigma$ as finite abstractions of \emph{discrete-time} stochastic systems $\widetilde\Sigma$ in~\eqref{discrete-version}. The abstraction algorithm is based on finite partitions of sets $\tilde X = \cup_i \mathsf X_i$, $\tilde U = \cup_i \mathsf U_i$, and $\tilde W = \cup_i \mathsf W_i$ and the selection of representative points $\bar \xi_i\in \mathsf X_i$, $\bar \nu_i\in \mathsf U_i$, and $\bar w_i\in \mathsf W_i$ as abstract states and inputs as formalized in the following definition.

\begin{definition}\label{finite system}	
	Given a discrete-time system $\widetilde \Sigma=(\tilde X,\tilde U, \tilde W,$ $\varsigma,\tilde f,\tilde Y_1,\tilde Y_2, \tilde h_1,\tilde h_2)$, its finite abstraction $\widehat\Sigma$ can be characterized as	
	\begin{equation}
	\label{eq:abs_tuple}
	\widehat\Sigma =(\hat X, \hat U,\hat W, \varsigma,\hat f,\hat Y_1,\hat Y_2,\hat h_1,\hat h_2),
	\end{equation}
	where $\hat X = \{\bar \xi_i,i=1,\ldots,n_{\tilde \xi}\}$, $\hat U = \{\bar \nu_i,i=1,\ldots,n_{\tilde \nu}\}$, and $\hat W = \{\bar w_i,i=1,\ldots,n_{\tilde w}\}$ are sets of selected representative points. Function $\hat f:\hat X\times\hat U\times\hat W\times \mathcal V_\varsigma\rightarrow\hat X$ is defined as
	\begin{equation}
	\label{eq:abs_dyn}
	\hat f(\hat \xi,\hat{\nu},\hat w,\varsigma) = \Pi_{\tilde\xi}(\tilde f(\hat \xi,\hat{\nu},\hat w,\varsigma)),	
	\end{equation}
	where $\Pi_{\tilde\xi}:\tilde X\rightarrow \hat X$ is a map that assigns to any $\tilde \xi\in \tilde X$, the representative point $\bar \xi\in\hat X$ of the corresponding partition set containing $\tilde \xi$.
	The output maps $\hat h_1$, $\hat h_2$ are the same as $\tilde h_1$, $\tilde h_2$ with their domain restricted to the finite state set $\hat X$ and the output sets $\hat Y_1$, $\hat Y_2$ are just the image of $\hat X$ under $\tilde h_1$, $\tilde h_2$.
	The initial state of $\widehat\Sigma$ is also selected according to $\hat \xi_0 := \Pi_{\tilde\xi}(\tilde \xi_0)$ with $\tilde \xi_0$ being the initial state of $\widetilde \Sigma$.
\end{definition}
The abstraction map $\Pi_{\tilde\xi}$ defined in \eqref{eq:abs_dyn}
satisfies the inequality
\begin{equation}\label{eq:Pi_delta}
\Vert \Pi_{\tilde\xi}(\tilde \xi)-\tilde \xi\Vert \leq \delta,~\quad \forall \tilde \xi\,\in \tilde X,
\end{equation}
where $\delta$ is the \emph{state} discretization parameter defined as $\delta:=\sup\{\|\tilde \xi-\tilde \xi'\|,\,\, \tilde \xi,\tilde \xi'\in \mathsf X_i,\,i=1,2,\ldots,n_{\tilde \xi}\}$.

\begin{remark}
	Note that to construct finite abstractions as in Definition~\ref{finite system}, we assume the state and input sets of the discrete-time system $\widetilde\Sigma$ are restricted to compact regions.
\end{remark}	

\section{Stochastic Storage and Simulation Functions}\label{sec:SSF}
In this section, we first define a notion of stochastic
storage functions (SStF) for ct-SCS with both internal and external signals. We then define a notion of stochastic simulation functions (SSF) for ct-SCS with only external signals. We utilize these two definitions to quantify the probabilistic closeness of interconnected \emph{continuous-time} stochastic systems and that of their \emph{discrete-time} (finite or infinite) abstractions.

\begin{definition}\label{Def_1a}
	Consider  a ct-SCS $\Sigma=(X,U,W,\mathcal U,\mathcal W,f,\sigma,Y_1,Y_2,h_1,h_2)$ and its (in)finite abstraction
	$\widehat\Sigma =(\hat X,\hat U,\hat W, \varsigma, \hat f,\hat Y_1,\hat Y_2,\hat h_1,\hat h_2)$. A function $S:X\times\hat X\to\mathbb R_{\ge0}$ is
	called a stochastic storage function (SStF) from  $\widehat\Sigma$ to $\Sigma$ if 
	\begin{itemize}
		\item  $\exists \alpha\in \mathcal{K}_{\infty}$ such that
		\begin{align}\label{eq:V_dec1}
		\forall x\in X,\forall \hat x\in\hat X, 	~~\alpha(\Vert h_1(x)-\hat h_1(\hat x)\Vert)\le S(x,\hat x),
		\end{align}
		\item $\forall k\in\mathbb N$, $\forall \xi := \xi(k\tau)\in X,\forall\hat \xi := \hat \xi(k)\in\hat X$, and $\forall\hat\nu:=\hat\nu(k)\in\hat U$, $\forall w := w(k\tau)\in W$, $ \forall\hat w:=\hat w(k)\in\hat W$, $\exists \nu:=\nu(k\tau)\in U$ such that
		\begin{align}\label{eq:V_dec}
		\EE& \Big[S(\xi((k+1)\tau),\hat{\xi}(k+1))\,\big|\,\xi,\hat{\xi},\nu,\hat \nu, w, \hat w\Big]\\\notag
		&\leq\kappa S(\xi,\hat \xi) +\rho_{\mathrm{ext}}(\Vert\hat\nu\Vert)+\psi
		+\begin{bmatrix}
		w-\hat w\\
		h_2(x)-\hat h_2(\hat x)
		\end{bmatrix}^T
		\underbrace{\begin{bmatrix}
			\bar X^{11}&\bar X^{12}\\
			\bar X^{21}&\bar X^{22}
			\end{bmatrix}}_{\bar X:=}\begin{bmatrix}
		w-\hat  w\\
		h_2(x)-\hat h_2(\hat x)
		\end{bmatrix}\!\!,
		\end{align}
		for some chosen sampling time $\tau \in \mathbb R_{> 0}$, $0<\kappa<1$, $\rho_{\mathrm{ext}}\in \mathcal{K}_{\infty}$, $\psi \in\R_{> 0}$, and a symmetric matrix $\bar X$ with conformal block partitions $\bar X^{ij}$, $i,j\in\{1,2\}$.
	\end{itemize}
\end{definition}

We call the control system $\widehat\Sigma$ a \emph{discrete-time} (in)finite abstraction of concrete (original) system $\Sigma$
if there exists an SStF $S$ from $\widehat\Sigma$ to $\Sigma$. Abstraction $\widehat{\Sigma}$ could be finite or infinite depending on cardinalities of sets $\hat X,\hat U, \hat W$. Since the above definition does not put any restriction on the state set of abstract systems, it can be also used to define a stochastic storage function from discrete-time system $\widetilde \Sigma$ presented in~\eqref{eq:dt-SCS} to $\Sigma$ (cf. the case study).

\begin{remark}
	Note that one can rewrite the left-hand side of~\eqref{eq:V_dec} using Dynkin's formula~\cite{dynkin1965markov} as
	\begin{align}\notag
	\EE& \Big[S(\xi((k+1)\tau),\hat{\xi}(k+1))\,\big|\,\xi(k\tau),\hat{\xi}(k), \nu(k\tau),\hat \nu(k), w(k\tau),\hat w(k)\Big]\\\notag
	&=\EE_{\varsigma} \Big[S(\xi(k\tau),\hat \xi(k+1))+\EE\Big[\int_{k\tau}^{(k+1)\tau}\mathcal{L}S(\xi(t),\hat \xi(k+1))\mathsf{d}t \Big]\,\big|\,\hat{\xi}(k),\hat \nu(k), \hat w(k)\Big]\!,\notag
	\end{align}
	where $\mathcal{L}S$ is the \emph{infinitesimal generator} of the stochastic process applying on the function $S$, and $\EE_{\varsigma}$ is the \emph{conditional expectation} acting only on the noise of the abstract system. The above Dynkin's formula is utilized later in Section~\ref{sec:constrcution_finite} to show the results of Theorem~\ref{Thm_3a}.
\end{remark}

Now, we write the above notion for the interconnected ct-SCS as the following definition.

\begin{definition}\label{Def_1b}
	Consider a ct-SCS $\Sigma=(X,U,\mathcal U,f,\sigma, Y, h)$ and its finite abstraction
	$\widehat\Sigma =(\hat X,\hat U, \varsigma, \hat f,\hat Y, \hat h)$ without internal signals. A function $V:X\times\hat X\to\mathbb R_{\ge0}$ is
	called a stochastic simulation function (SSF) from  $\widehat\Sigma$ to $\Sigma$ if 
	\begin{itemize}
		\item  $\exists \alpha\in \mathcal{K}_{\infty}$ such that $\forall x\in X,\forall \hat x\in\hat X,$ one has
		\begin{align}\label{eq:V_dec1b}
		\alpha(\Vert h(x)-\hat h(\hat x)\Vert)\le V(x,\hat x),
		\end{align}
		\item $\forall k\in\mathbb N$, $\forall \xi := \xi(k\tau)\in X,\forall\hat \xi := \hat \xi(k)\in\hat X$, and $\forall\hat\nu:=\hat\nu(k)\in\hat U$, $\exists \nu:=\nu(k\tau)\in U$ such that
		\begin{align}\label{eq:V_decb}
		\EE \Big[V(\xi((k+1)\tau),\hat{\xi}(k+1))\,\big|\,\xi,\hat{\xi},\nu,\hat \nu\Big]\leq\kappa V(\xi,\hat \xi) +\rho_{\mathrm{ext}}(\Vert\hat\nu\Vert)+\psi,
		\end{align}
		for some chosen sampling time $\tau \in \mathbb R_{> 0}$, $0<\kappa<1$, $\rho_{\mathrm{ext}} \in \mathcal{K}_{\infty}$, and $\psi \in\mathbb R_{> 0}$.
	\end{itemize}
\end{definition}

The next theorem is borrowed from~\cite[Theorem 3.3]{lavaei2017compositional} and shows how SSF can be useful in providing the probabilistic closeness between output trajectories of original interconnected  \emph{continuous-time} stochastic systems and that of their \emph{discrete-time} (finite or infinite) abstractions.

\begin{theorem}\label{Thm_1a}
	Let
	$	\Sigma=(X,U,\mathcal U,f,\sigma ,Y, h)$ be a ct-SCS and
	$\widehat\Sigma =(\hat X,\hat U,\varsigma,\hat f,\hat Y,\hat h)$ its \emph{discrete-time} abstraction.
	Suppose $V$ is an SSF from $\widehat\Sigma$ to $\Sigma$. For any input trajectory $\hat\nu(\cdot)\in\mathcal{\hat U}$ that preserves Markov property for the closed-loop $\widehat\Sigma$, and for any random variables $a$ and $\hat a$ as initial states of the ct-SCS and its \emph{discrete-time} abstraction, there exists an input trajectory $\nu(\cdot)\in\mathcal{U}$ of $\Sigma$ such that the following inequality holds over the finite-time horizon $T_d$:	
	\begin{align}\label{Eq_25}
	\mathds{P}&\left\{\sup_{0\leq k\leq T_d}\Vert \zeta_{a\nu}(k\tau)-\hat \zeta_{\hat a \hat\nu}(k)\Vert\geq\varepsilon\,|\,a,\hat a\right\}\\\notag
	&\leq
	\begin{cases}
	1-(1-\frac{V(a,\hat a)}{\alpha\left(\varepsilon\right)})(1-\frac{\hat\psi}{\alpha\left(\varepsilon\right)})^{T_d}, & \quad\quad\text{if}~\alpha\left(\varepsilon\right)\geq\frac{\hat\psi}{\kappa},\\
	(\frac{V(a,\hat a)}{\alpha\left(\varepsilon\right)})(1-\kappa)^{T_d}+(\frac{\hat\psi}{\kappa\alpha\left(\varepsilon\right)})(1-(1-\kappa)^{T_d}), & \quad\quad\text{if}~\alpha\left(\varepsilon\right)<\frac{\hat\psi}{\kappa},
	\end{cases}
	\end{align}
	where $\hat\psi>0$ satisfies  $\hat\psi\geq \rho_{\mathrm{ext}}(\Vert\hat \nu\Vert_{\infty})+\psi$.
\end{theorem}

\section{Compositional Abstractions for Interconnected Systems}
\label{Compositionality Results}
In this section, we analyze networks of stochastic control subsystems, $i\in \{1,\dots,N\},$
\begin{equation}
\label{eq:network}
\Sigma_i=(X_i,U_i,W_i,\mathcal U_i,\mathcal W_i,f_i,\sigma_i,Y_{1_i}, Y_{2_i},h_{1_i},h_{2_i}),
\end{equation}
and discuss how to construct their finite abstractions together with an SSF based on corresponding SStF of their subsystems. 

\subsection{Interconnected Stochastic Control Systems}\label{Interconnected SCS}
We first formally define the interconnected stochastic control systems.

\begin{definition}\label{Concrete interconnection}
	Consider $N\!\in\!\N_{\geq1}$ stochastic control subsystems $\Sigma_i\!=\!(X_i,U_i,W_i,\mathcal U_i,\mathcal W_i,f_i,\sigma_i,Y_{1_i}, Y_{2_i},\!h_{1_i},\!h_{2_i}\!)$, $i\in \{1,\dots,N\}$,
	and a matrix $M$ defining the coupling between these subsystems. We require the condition $M\prod_{i=1}^N Y_{2i} \subseteq \prod_{i=1}^N W_{i}$ to establish a well-posed interconnection. The interconnection of  $\Sigma_i$,
	$\forall i\in \{1,\ldots,N\}$,
	is the ct-SCS $\Sigma=(X,U,\mathcal U,f,\sigma,Y, h)$, denoted by
	$\mathcal{I}(\Sigma_1,\ldots,\Sigma_N)$, such that $X:=\prod_{i=1}^{N}X_i$,  $U:=\prod_{i=1}^{N}U_i$, $f:=\prod_{i=1}^{N}f_{i}$,
	$\sigma:=[\sigma_1(x_1);\cdots;\sigma_N(x_N)]$, $Y:=\prod_{i=1}^{N}Y_{1i}$, and $h=\prod_{i=1}^{N}h_{1i}$, with the internal inputs constrained according to:
	\begin{align}\notag
	\intcc{w_{1};\cdots;w_{N}}=M\intcc{h_{21}(x_1);\cdots;h_{2N}(x_N)}.
	\end{align}
\end{definition}

\begin{remark}
	Note that we do not have any restrictions on the interconnected matrix $M$ and its entries can take any values depending on the forms of interconnection topologies.	
\end{remark}

\subsection{Compositional Abstractions of Interconnected Systems}\label{sec:constrcution_infinite}

We consider $\Sigma_i\!=\!(X_i,U_i,W_i,\mathcal U_i,\mathcal W_i,f_i,\\\sigma_i,Y_{1_i}, Y_{2_i},h_{1_i},h_{2_i})$ as an original ct-SCS and $\widehat{\Sigma}_i$ as its discrete-time finite abstraction given by the tuple 
$\widehat\Sigma_i=(\hat X_i,\hat U_i,\hat W_i,\varsigma_i,\hat f_i,\hat Y_{1_i}, \hat Y_{2_i},\hat h_{1_i},\hat h_{2_i})$. We also assume that there exist an SStF $S_i$ from $\widehat\Sigma_i$ to $\Sigma_{i}$ with the corresponding functions, constants, and matrices denoted by $\alpha_i$, $\rho_{\text{ext}i}$, $\kappa_i$, $\psi_i$, $\bar X_i$, $\bar X_i^{11}$, $\bar X_i^{12}$, $\bar X_i^{21}$, and $\bar X_i^{22}$. In the next theorem, we quantify the error between the interconnection of \emph{continuous-time} stochastic subsystems and that of their \emph{discrete-time} abstractions in a compositional fashion.

\begin{theorem}\label{Thm: Comp1}
	Consider an interconnected stochastic control system
	$\Sigma=\mathcal{I}(\Sigma_1,\ldots,\Sigma_N)$ induced by $N\in{\N_{\geq1}}$ stochastic
	control subsystems~$\Sigma_i$ and the coupling matrix $M$. Let each subsystem $\Sigma_i$ admit an abstraction $\widehat \Sigma_i$ with the corresponding SStF $S_i$.
	Then
	\begin{equation}
	\label{Comp: Simulation Function1}
	V(x,\hat x)\Let\sum_{i=1}^N\mu_iS_i(x_i,\hat x_i),
	\end{equation}
	is a stochastic simulation function from the interconnected system
	$\widehat \Sigma=\mathcal{I}(\widehat \Sigma_1,\ldots,\widehat\Sigma_N)$, with coupling matrix $\hat M$, to $\Sigma=\mathcal{I}(\Sigma_1,\ldots,\Sigma_N)$
	if there exist $\mu_{i}>0$, $i\in\{1,\ldots,N\}$, and
	\begin{align}\label{Con_1a}
	\begin{bmatrix}
	M\\\mathds{I}_{\tilde q}
	\end{bmatrix}^T &\bar X_{cmp}\begin{bmatrix}
	M\\\mathds{I}_{\tilde q}
	\end{bmatrix}\preceq0,
	\\\label{Con_2a}
	M&=\hat M,\\\label{Con111}
	\hat M\prod_{i=1}^N \hat Y_{2i} &\subseteq \prod_{i=1}^N \hat W_{i},
	\end{align}
	where
	\begin{equation}
	\bar X_{cmp}:=\begin{bmatrix}
	\mu_1\bar X_1^{11}&&&\mu_1\bar X_1^{12}&&\\
	&\ddots&&&\ddots&\\
	&&\mu_N\bar X_N^{11}&&&\mu_N\bar X_N^{12}\\
	\mu_1\bar X_1^{21}&&&\mu_1\bar X_1^{22}&&\\
	&\ddots&&&\ddots&\\
	&&\mu_N\bar X_N^{21}&&&\mu_N\bar X_N^{22}
	\end{bmatrix}\!\!,\\\label{Def_3a}
	\end{equation}
	and $\tilde q=\sum_{i=1}^Nq_{2i}$ with $q_{2i}$ being dimensions of the internal output of subsystems $\Sigma_i$.
\end{theorem}

\begin{remark}\label{compositionality remark}
	Condition \eqref{Con_1a} is similar to the LMI discussed in \cite{2016Murat} as a compositional stability condition based on the dissipativity theory. It is shown in \cite{2016Murat} that this condition holds independently of the number of subsystems in many physical applications with particular interconnection structures, e.g., skew symmetric.
\end{remark}

\section{Construction of Stochastic Storage Functions for a Class of Systems}
\label{sec:constrcution_finite}

In this section, we focus on a special class of continuous-time stochastic affine systems and impose conditions enabling us to establish an SStF from its finite abstraction $\widehat{\Sigma}$ to $\Sigma$. The model of the system is given by
\begin{align}\label{Eq_5a}
\Sigma:\left\{\hspace{-1mm}\begin{array}{l}\mathsf{d}\xi(t)=(A\xi(t)+B\nu(t)+Dw(t)+\mathbf{b})\mathsf{d}t+G\mathsf{d}W_t,\\
\zeta_1(t)=C_1\xi(t),\\
\zeta_2(t)=C_2\xi(t),\\
\end{array}\right.
\end{align}
where $A\in\mathbb R^{n\times n}, B\in\mathbb R^{n\times m}, D\in\mathbb R^{n\times p}, C_1\in\mathbb R^{q_1\times n}, C_2\in\mathbb R^{q_2\times n}$, $G\in\mathbb R^{n}$, and $\mathbf{b}\in\mathbb R^{n}$. We employ the tuple
\begin{align}\notag
\Sigma=(A,B,C_1, C_2,D,G,\mathbf{b}),
\end{align}
to refer to the class of stochastic affine systems in~\eqref{Eq_5a}. The \emph{time-discretized} version of $\Sigma$ is proposed as
\begin{align}\label{time-discretized}
\widetilde\Sigma:\left\{\hspace{-1mm}\begin{array}{l}\tilde \xi(k+1)=\tilde \xi(k)+\tilde \nu(k)+\tilde D\tilde w(k)+\tilde R\varsigma(k),\\
\tilde \zeta_1(k)=\tilde C_1 \tilde \xi(k),\\
\tilde \zeta_2(k)=\tilde C_2 \tilde \xi(k),\\
\end{array}\right.
\quad k\in\mathbb N,
\end{align}
where $\tilde D$ and $\tilde R$ are matrices chosen arbitrarily, and $\tilde C_1 = C_1P$, $\tilde C_2 = C_2P$ with $P$ as chosen in~\eqref{Eq_7a} (cf. Theorem~\ref{Thm_3a}). Our main target here is to employ $\widetilde\Sigma$ as the discrete-time version of $\Sigma$ in order to establish an SStF from $\widehat\Sigma$ to $ \Sigma$ through $\widetilde\Sigma$ while quantifying the \emph{best approximation error}. Later, in Remark~\ref{Remark: Non-stochastic}, we show that $\tilde R = \mathbf{0}_n$ and $\tilde D = \mathbf{0}_{n\times p}$ result in the least approximation error in our settings. Now, we describe the \emph{finite} abstraction of $\widetilde\Sigma$ as
\begin{align}\label{Eq:1}
\widehat\Sigma:\left\{\hspace{-1mm}\begin{array}{l}\hat \xi(k+1)=\Pi_{\tilde\xi}(\hat \xi(k)+\hat \nu(k)+\tilde D\hat w(k)+\tilde R \varsigma(k)),\\
\hat \zeta_1(k)=\hat C_1\hat \xi(k),\\
\hat \zeta_2(k)=\hat C_2\hat \xi(k),\\
\end{array}\right.
\quad k\in\mathbb N,
\end{align}
where map $\Pi_{\tilde\xi}: \tilde X \rightarrow \hat X$ satisfies the inequality~\eqref{eq:Pi_delta}. Now we candidate the following quadratic stochastic storage function
\begin{align}\label{Eq_7a}
S(x,\hat x)=(x-P\hat x)^T \mathcal {\bar M}(x-P\hat x),
\end{align}
where $P$ is a square matrix and $\mathcal {\bar M}$ is a positive-definite matrix of an appropriate dimension. In order to show that $S$ in \eqref{Eq_7a} is an SStF from $\widehat\Sigma$ to $\Sigma$, we need the following key assumptions over $\Sigma$. 

\begin{assumption}\label{ASM: 1}
	Assume that there exists a \emph{concave} function $\gamma\in\mathcal{K}_{\infty}$  such that $S$ satisfies
	\begin{align}\label{Eq65}
	S(x,x')-S(x,x'')\leq \gamma(\Vert x'-x''\Vert),
	\end{align}
	for any $ x,x',x''\in X$.
\end{assumption}

Note that Assumption \ref{ASM: 1} is always fulfilled for the function $S$ in \eqref{Eq_7a} as long as it is restricted to a compact subset of $X \times X$.

\begin{assumption}\label{ASM: 2}
	Let $\Sigma=(A,B,C_1,C_2,D,G,\mathbf{b})$. Assume that for some constant $\tilde \kappa\in\R_{> 0}$, there exist matrices $\ \mathcal {\bar M}\succ0$, $K$, $Q$ and $H$ of appropriate dimensions such that the following matrix (in)equalities hold:
	\begin{IEEEeqnarray}{rCl}\IEEEyesnumber\label{Con_LMI}
		\IEEEyessubnumber\label{Con_1} (A+BK)^T\mathcal {\bar M}+\mathcal {\bar M}(A+BK)&\leq& -\tilde \kappa \mathcal {\bar M},\\
		\IEEEyessubnumber\label{Con_2}  BQ&=&AP,\\
		\IEEEyessubnumber\label{Con_3}  D&=&BH.
	\end{IEEEeqnarray}
\end{assumption}

Note that stabilizability of the pair $(A,B)$ is necessary and sufficient
to satisfy condition~\eqref{Con_1}. Moreover, there exist matrices $Q$ and $H$ satisfying conditions~\eqref{Con_2} and~\eqref{Con_3} if and only if $\text{im}~AP \subseteq \text{im}~B$ and $\text{im}~D \subseteq \text{im}~B$, respectively.

\begin{assumption}\label{ASM: 3}
	Let $\Sigma=(A,B,C_1,C_2, D,G,\mathbf{b})$. Assume that for some constants
	$\pi>0$ and $0<\bar \kappa < 1 - e^{-\tilde \kappa \tau}$  with a sampling time $\tau$, there exist matrices $\bar X^{11}$, $\bar X^{12}$, $\bar X^{21}$, and $\bar X^{22}$ of appropriate dimensions such that
	\begin{align}\label{Eq_8a}
	&\begin{bmatrix}
	\pi e^{-\tilde \kappa \tau}\tau B^T\mathcal {\bar M}B&&0 \\
	0&& \pi e^{-\tilde \kappa \tau}\tau D^T \mathcal {\bar M}D\\
	\end{bmatrix}\preceq\begin{bmatrix}
	\bar \kappa \mathcal {\bar M} + C_2^T\bar X^{22}C_2 & C_2^T\bar X^{21}\\
	\bar X^{12}C_2 & \bar X^{11}\\
	\end{bmatrix}\!\!,
	\end{align}
	where $\mathcal {\bar M}\succ0$ is the matrix appeared in~\eqref{Con_1}.
\end{assumption}

\begin{remark}
	Note that in Assumption~\ref{ASM: 3}, matrices $B, D, C_2$ are those in the system dynamics, constant and matrix $\tilde \kappa, \mathcal {\bar M}$ are the same as those satisfying the condition~\eqref{Con_1}, and constants and matrices $\pi, \tau, \bar \kappa, \bar X^{11}, \bar X^{12}, \bar X^{21}, \bar X^{22}$ are our decision variables to be designed. One can readily satisfy this assumption via semi-definite programing toolboxes and then check the compositionality condition~\eqref{Con_1a} with obtained conformal block partitions $\bar X^{ij}$, $i,j\in\{1,2\}$ of subsystems (cf. the case study).  
\end{remark}

Now we provide another main result of the paper showing that under which conditions $S$ in \eqref{Eq_7a} is an SStF from $\widehat \Sigma$ to $\Sigma$. 
\begin{theorem}\label{Thm_3a}
	Let $\Sigma=(A,B,C_1, C_2,D,G,\mathbf{b})$ and $\widehat \Sigma$ be its \emph{finite} MDP with discretization parameter $\delta$. Suppose Assumptions~\ref{ASM: 1},~\ref{ASM: 2} and~\ref{ASM: 3} hold, $\hat C_1 = \tilde C_1 = C_1P$, and $\hat C_2 = \tilde C_2 = C_2P$. Then the quadratic function $S$ in~\eqref{Eq_7a} is an SStF from $\widehat \Sigma$ to $\Sigma$.
\end{theorem}

The functions and constants
$\alpha,\rho_{\mathrm{ext}}\in \mathcal{K}_{\infty}$, $0<\kappa<1$, and $\psi \in\R_{> 0}$ in Definition~\ref{Def_1a} associated with $S$ in~\eqref{Eq_7a} are computed as
\begin{align}\notag
\alpha(s)&:=\frac{\lambda_{\min}(\mathcal {\bar M})}{\lambda_{\max}(C_1^TC_1)}s^2, \quad \forall s\in\R_{\geq0},\\\notag
\kappa &:= \bar\kappa+e^{-\tilde \kappa \tau},\\\notag
\rho_{\mathrm{ext}}(s) &:= \gamma((1+\frac{1}{\bar\eta})(1+\bar\eta')(1+\bar\eta'') s), \quad \forall s\in\R_{\geq0},\\\notag
\psi &:=e^{-\bar \kappa \tau}\tau (G^T\mathcal {\bar M}G+\pi\Vert \sqrt{\mathcal {\bar M}}\mathbf{b}\Vert^2)+\gamma((1+\bar\eta)\delta)+\gamma((1+\frac{1}{\bar\eta})(1+\frac{1}{\bar\eta'})\sqrt{\text{Tr}(\tilde R^T \tilde R)})\\\notag
&\quad\quad\!\!\!\!+\gamma((1+\frac{1}{\bar\eta})(1+\bar\eta')(1+\frac{1}{\bar\eta ''}) \,\Vert \tilde D \,\Vert\, \Vert \hat w\Vert),
\end{align}
where $\bar\eta,\bar\eta', \bar\eta''>0$ are some positive constants chosen arbitrarily.

\begin{remark}\label{Remark: Non-stochastic}
	Note that for the discrete-time system $\widetilde \Sigma$ in \eqref{time-discretized}, $\rho_{\mathrm{ext}},$ and $\psi$ defined above reduce to
	\begin{align}\notag
	\rho_{\mathrm{ext}}(s) :=~&\gamma((1+\bar\eta)(1+\bar\eta')s),\quad \forall s\in\R_{\geq0},\\\notag
	\psi :=~&e^{-\tilde \kappa \tau}\tau (G^T\mathcal {\bar M}G+\pi\Vert \sqrt{\mathcal {\bar M}}\mathbf{b}\Vert^2)+\gamma((1+\frac{1}{\bar\eta})\sqrt{\text{Tr}(\tilde R^T \tilde R)})+\gamma((1+\bar\eta)(1+\frac{1}{\bar\eta '}) \,\Vert \tilde D \,\Vert\, \Vert \hat w\Vert).
	\end{align}
	Moreover, if the abstraction $\widetilde \Sigma$ is \emph{non-stochastic} (i.e., $\tilde R = \mathbf{0}_n$) with $\tilde D = \mathbf{0}_{n\times p}$,
	then
	\begin{align}\notag
	\rho_{\mathrm{ext}}(s) :=\gamma(s), \forall s\in\R_{\geq0},\quad\psi :=e^{-\tilde \kappa \tau}\tau (G^T\mathcal {\bar M}G+\pi\Vert \sqrt{\mathcal {\bar M}}\mathbf{b}\Vert^2).
	\end{align}
	This simply means if the concrete system satisfies some stability property (cf.~\eqref{Con_1}), it is better to pick non-stochastic discrete-time system rather than stochastic ones since the non-stochastic systems provide smaller approximation errors (cf. the case study). 
\end{remark}

Note that $\tilde D = \mathbf{0}_{n\times p}$ (\emph{i.e.,} not having any internal input in the abstract systems in \eqref{Eq:1}) will result in less approximation errors. In fact, a smart choice of the interface map~\eqref{Eq_405a} in Appendix
still ensures that the output trajectories of abstract systems follow those of the original ones with a quantified probabilistic error bound which gets smaller if $\tilde D = \mathbf{0}_{n\times p}$.

\section{Case Study}\label{case study}

\begin{figure}
	\centering
	\includegraphics[width=5.5cm]{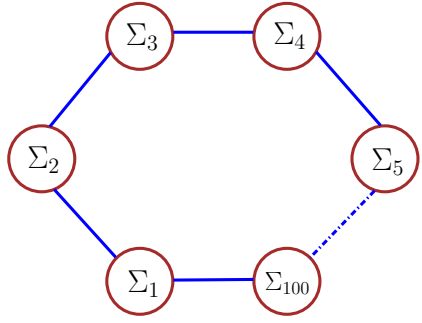}
	\caption{A circular building in a network of $100$ rooms.}
	\label{Case_Studies}
\end{figure}

To illustrate the effectiveness of the proposed results, we apply our approaches to a temperature regulation in a circular network containing $100$ rooms and construct compositionally a discrete-time system
from its original continuous-time dynamic. We then employ the constructed discrete-time abstractions as substitutes to compositionally synthesize policies regulating the temperature of each room in a comfort zone.

Consider a network of $n=100$ rooms each equipped with a heater and connected circularly as depicted in Figure~\ref{Case_Studies}.
The model of this case study is adapted from~\cite{girard2016safety}
by including stochasticity in
the model. The evolution of the temperature $T(\cdot)$ can be presented by the following interconnected stochastic differential equation
\begin{align}\label{Case study}
\Sigma:\left\{\hspace{-1mm}\begin{array}{l}{\mathsf{d}T(t)}=(A{T}(t)+\theta T_{h}\nu(t)+ \beta T_{E})\,\mathsf{d}t+G\mathsf{d}\mathbb W_t,\\
\zeta(t)={T}(t),\end{array}\right.
\end{align}
where $A$ is a matrix with diagonal elements $\bar a_{ii}=-2\eta-\beta-\theta\nu_i(t)$, $i\in\{1,\ldots,n\}$, off-diagonal elements $\bar a_{i,i+1}=\bar a_{i+1,i}=\bar a_{1,n}=\bar a_{n,1}=\eta$, $i\in \{1,\ldots,n-1\}$, and all other elements are identically zero, and $G = 0.5\mathds{I}_n$.
Parameters $\eta = 0.05$, $\beta = 0.005$, and $\theta = 0.01$ are conduction factors, respectively, between the rooms $i \pm 1$ and $i$, the external environment and the room $i$, and the heater and the room $i$. Moreover, $T_E=[T_{e_1};\ldots;T_{e_n}]$, $\nu(t)=[\nu_1(t);\ldots;\nu_n(t)]$, and $ T(t)=[T_1(t);\ldots;T_n(t)]$, where $T_i(t)$ is taking values in the set $[20,21]$, for all $i\in\{1,\ldots,n\}$. Outside temperatures are the same for all rooms: $T_{ei}=-1\,^\circ C$, $\forall i\in\{1,\ldots,n\}$, and the heater temperature is $T_h=50\,^\circ C$.
Now by considering the individual rooms as $\Sigma_i$ described by
\begin{align}\notag
\Sigma_i:\left\{\hspace{-1mm}\begin{array}{l}{\mathsf{d}T_i(t)}=(\bar a_{ii}{T_i}(t)+\theta T_{h}\nu_i(t)+\eta w_i(t)+\beta T_{e_i})\,\mathsf{d}t+0.5\mathsf{d}\mathbb W_{t_i},\\
\zeta_{1_i}(t)=T_{i}(t),\\
\zeta_{2_i}(t)=T_{i}(t),\end{array}\right.
\end{align}
one can readily verify that $\Sigma=\mathcal{I}(\Sigma_1,\ldots,\Sigma_N)$ where the coupling matrix $M$ is such that  $m_{i,i+1}=m_{i+1,i}=m_{1,n}=m_{n,1}=1$, $i\in \{1,\ldots,n-1\}$, and all other elements are identically zero. The discretized version of $\Sigma_i$ is proposed by
\begin{align}\notag
\widetilde\Sigma_i:\left\{\hspace{-1mm}\begin{array}{l}\tilde T_i(k+1)=\tilde T_i(k)+\tilde\nu_i(k),\\
\tilde \zeta_{1_i}(k)=\tilde T_i(k),\\
\tilde \zeta_{2_i}(k)=\tilde T_i(k),\\
\end{array}\right.
\quad\quad k\in\mathbb N.
\end{align}
As discussed in Remark~\ref{Remark: Non-stochastic}, we consider here $\tilde R_i = \tilde D_i = 0$ to have the least constants  $\psi_i$
for each $S_i$ (resulting in the least probabilistic error). Then, one can readily verify that conditions~\eqref{Con_1}-\eqref{Con_3} are satisfied by $\mathcal {\bar M}_i = 1, P_i = 1, Q_i = -0.21, H_i = 0.1$. Condition~\eqref{Eq_8a} is also satisfied with $\tau_i = 0.1, \pi_i = 1, \bar \kappa_i = 0.499, \bar X^{11}_i=e^{-\tilde \kappa_i \tau_i}\tau_i \eta^2$, $\bar X^{22}_i=-\pi_ie^{-\tilde \kappa_i \tau_i}\tau_i \theta^2 T_{h}^2$, $\bar X^{12}_i=\bar X^{21}_i=0$. Therefore, $S_i(T_i(k\tau),\tilde T_i(k))=(T_i(k\tau)-\tilde T_i(k))^2$ is an SStF from $\widetilde\Sigma_i$ to $\Sigma_i$ satisfying the condition~\eqref{eq:V_dec1} with $\alpha_i(s)=s^2, \forall s\in \mathbb R_{\ge0}$ and the condition~\eqref{eq:V_dec} with $\kappa_i=0.5$, $\rho_{\mathrm{ext}i}(s)=2s$, $\forall s\in \mathbb R_{\ge0}$, $\psi_i =  1.17 \times 10^{-10}$, and

\begin{align}\label{Eq_22}
\bar X_i=\begin{bmatrix} e^{-\tilde \kappa_i \tau_i}\tau_i\eta^2 & 0  \\ 0  &  -\pi_ie^{-\tilde \kappa_i \tau_i}\tau_i \theta^2 T_{h}^2 \end{bmatrix}\!\!.
\end{align}

Now we look at $\widetilde\Sigma=\mathcal{I}(\widetilde\Sigma_1,\ldots,\widetilde\Sigma_N)$
with a coupling matrix $\tilde M$ satisfying the condition \eqref{Con_2a} as $\tilde M = M$.
Choosing $\mu_1=\cdots=\mu_N=1$ and using $\bar  X_i$ in (\ref{Eq_22}), matrix $\bar X_{cmp}$ in \eqref{Def_3a} reduces to
$$
\bar X_{cmp}=\begin{bmatrix} e^{-\tilde \kappa_i \tau_i}\tau_i \eta^2\mathds{I}_n & 0 \\ 0 & -\pi_ie^{-\tilde \kappa_i \tau_i}\tau_i \theta^2 T_{h}^2\mathds{I}_n \end{bmatrix}\!,
$$
and accordingly the condition \eqref{Con_1a} reduces to
\begin{align}\notag
\begin{bmatrix} M \\ \mathds{I}_n \end{bmatrix}^T\!\bar X_{cmp}\begin{bmatrix} M \\ \mathds{I}_n \end{bmatrix}=e^{-\tilde \kappa_i \tau_i}\tau_i \eta^2M^TM-\pi_ie^{-\tilde \kappa_i \tau_i}\tau_i \theta^2 T_{h}^2\mathds{I}_n \preceq 0,
\end{align}
without requiring any restrictions on the  number or gains of subsystems.
We used $M=M^T$\!, and $4e^{-\tilde \kappa_i \tau_i}\tau_i \eta^2-\pi_ie^{-\tilde \kappa_i \tau_i}\tau_i \theta^2 T_{h}^2\preceq 0$ by employing Gershgorin circle theorem \cite{bell1965gershgorin} to show the above LMI.
Hence, $V(T(k\tau),\tilde T(k))=\sum_{i=1}^{100} (T_i(k\tau)-\tilde T_i(k))^2$ is an SSF from  $\widetilde\Sigma$ to $\Sigma$ satisfying conditions \eqref{eq:V_dec1b} and \eqref{eq:V_decb}  with $\alpha(s)=s^2, \forall s\in \mathbb R_{\ge0}$, $\kappa=0.5$, $\rho_{\mathrm{ext}}(s)=20s, \forall s\in \mathbb R_{\ge0}$, and $\psi = 1.17 \times 10^{-8}$.

By taking initial states of $\Sigma$ and $ \widetilde \Sigma$
as $20.5\mathds{1}_{100}$, and employing Theorem~\ref{Thm_1a}, one can guarantee that the distance between outputs of $\Sigma$ and $\widetilde \Sigma$ will not exceed $\varepsilon = 0.5$ during the time horizon $T_d=12$ with a probability at least $91\%$, \emph{i.e.,}
\begin{equation}\notag
\mathds{P}(\Vert \zeta(k\tau)-\tilde \zeta(k)\Vert\le 0.5,\,\, \forall k\in[0,12])\ge 0.91\,.
\end{equation}

\begin{figure}
	\centering
	\includegraphics[width=9cm]{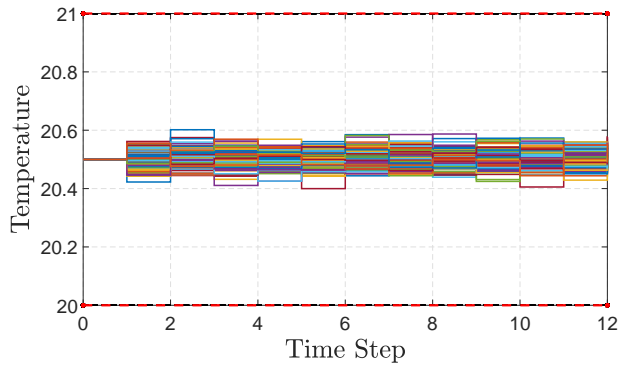}
	\caption{Closed-loop state trajectories of a representative room with different noise realizations in a network of $100$ rooms, for $T_d = 12$.}
	\label{Simulation}
\end{figure}

\begin{figure}
	\centering
	\includegraphics[width=9cm]{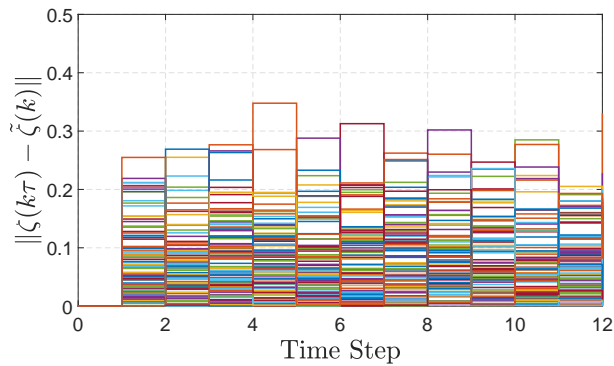}
	\caption{Several realizations of the norm of the error between the outputs of $\Sigma$ and of $\widetilde \Sigma$, \emph{i.e.,} $\Vert \zeta(k\tau)-\tilde \zeta(k)\Vert$, for $T_d=12$.}
	\label{Simulation1}
\end{figure}

We now synthesize a controller for $\Sigma$ via its \emph{discrete-time}
system $\widetilde \Sigma$ such that the controller keeps the temperature of each room in the comfort zone $[20,21]$. The idea here is to design a local controller for the abstract subsystem $\widetilde \Sigma_i$, and then refine it back to the subsystem $\Sigma_i$. Consequently, controller for the interconnected system $\Sigma$ would be a vector such that each of its components is the controller for subsystems $\Sigma_i$. We employ the software tool \texttt{SCOTS} \cite{rungger2016scots} to synthesize controllers for $\widetilde\Sigma_i$ maintaining the temperature of each room in the safe set $[20,21]$. Closed-loop state trajectories of a representative room with different noise realizations in a network of $100$ rooms are illustrated in Figure~\ref{Simulation}. Furthermore, several realizations of the norm of the error between outputs of $\Sigma$ and $\widetilde \Sigma$ are illustrated in Figure~\ref{Simulation1}.


\section{Discussion}
In this paper, we provided a compositional scheme for constructing finite MDPs of continuous-time stochastic control systems. We first defined notions of
stochastic storage and simulation functions between original \emph{continuous-time} stochastic systems and their \emph{discrete-time} (finite or infinite) abstractions with and without internal signals. We then leveraged dissipativity-type compositional conditions for the compositional quantification of the distance between the interconnection of concrete continuous-time stochastic control systems and their discrete-time (in)finite abstractions. We focused on a particular class of stochastic affine systems and constructed their finite abstractions together with their corresponding stochastic storage functions. We finally illustrated the effectiveness of our proposed results by applying them to a physical case study.

\section{Acknowledgment}
The authors would like to thank Abolfazl Lavaei for the fruitful discussions and helpful comments.

\bibliographystyle{alpha}
\bibliography{biblio}

\section{Appendix}
\begin{proof}\textbf{(Theorem~\ref{Thm: Comp1})}
	We first show that the SSF $V$ in~\eqref{Comp: Simulation Function1} satisfies the inequality \eqref{eq:V_dec1b} for some $\mathcal{K}_\infty$ function $\alpha$. For any $x=[{x_1;\ldots; x_N}]\in X$ and  $\hat x=[{\hat x_1;\ldots;\hat x_N}]\in \hat X$, one gets:
	\begin{align}\notag
	\Vert h(x)-\hat h(\hat x) \Vert&=\Vert [h_{11}(x_1);\ldots;h_{1N}(x_N)]-[\hat h_{11}(\hat x_1);\ldots;\hat h_{1N}(\hat x_N)]\Vert\\\notag
	&\le\sum_{i=1}^N \Vert  h_{1i}(x_i)-\hat h_{1i}(\hat x_i) \Vert~\le \sum_{i=1}^N \alpha_{i}^{-1}(S_i( x_i, \hat x_i))\le \bar\alpha(V(x,\hat x)),
	\end{align}
	with the function $\bar\alpha:\mathbb R_{\ge 0}\rightarrow\mathbb R_{\ge 0}$ defined for all $s\in\mathbb R_{\ge 0}$ as
	
	\begin{center}
		$\bar\alpha(s) \Let \max\left\{\sum_{i=1}^N\alpha_{i}^{-1}(s_i)\,\,\big|\, s_i  {\ge 0},\,\,\sum_{i=1}^N \mu_i s_i=s\right\}\!. $
	\end{center}
	By taking the $\mathcal{K}_\infty$ function $\alpha(s):=\bar\alpha^{-1}(s)$, $\forall s\in\R_{\ge0}$, one acquires
	$$\alpha(\Vert h(x)-\hat h(\hat x)\Vert)\le V( x, \hat x),$$
	fulfilling the inequality \eqref{eq:V_dec1b}.
	
	We continue with showing~\eqref{eq:V_decb}, as well. One can obtain the chain of inequalities in \eqref{Eq_4a}
	using conditions \eqref{Con_1a} and \eqref{Con_2a} and by defining $\kappa(\cdot),\psi,\rho_{\mathrm{ext}}(\cdot)$ as
	\begin{align*}
	\kappa s&\Let \max\Big\{\sum_{i=1}^N\mu_i\kappa_i s_i\,\,\big|\, s_i  {\ge 0},\,\,\sum_{i=1}^N \mu_i s_i=s\Big\},\\
	\rho_{\mathrm{ext}}(s) &\Let \max\Big\{\sum_{i=1}^N\mu_i\rho_{\mathrm{ext}i}(s_i)\,\,\big|\, s_i  {\ge 0},\,\,\|\intcc{s_1;\ldots;s_N}\| = s\Big\},\\
	\psi&\Let\sum_{i=1}^N\mu_i\psi_i.
	\end{align*}
	Hence one can conclude that $V$ is an SSF from $\widehat \Sigma$ to $\Sigma$.
	
	\begin{figure*}[ht!]
		\rule{\textwidth}{0.1pt}
		\begin{align}\notag
		\mathds{E}&\Big[V(\xi((k+1)\tau),\hat{\xi}(k+1))\,\big|\,\xi,\hat{\xi},\nu,\hat \nu\Big]=\mathbb{E}\Big[\sum_{i=1}^N\mu_iS_i(\xi_i((k+1)\tau),\hat{\xi}_i(k+1))\,\big|\,\xi,\hat{\xi},\nu,\hat \nu\Big]\\\notag
		&=\sum_{i=1}^N\mu_i\mathbb{E}\Big[S_i(\xi_i((k+1)\tau),\hat{\xi}_i(k+1))\,\big|\,\xi,\hat{\xi},\nu,\hat \nu\Big]= \sum_{i=1}^N\mu_i\mathbb{E}\Big[S_i(\xi_i((k+1)\tau),\hat{\xi}_i(k+1))\,\big|\,\xi_i,\hat{\xi_i},\nu_i,\hat \nu_i\Big]\\\notag
		&\leq\sum_{i=1}^N\mu_i\bigg(\kappa_iS_i( x_i,\hat x_i)+\rho_{\mathrm{ext}i}(\Vert \hat \nu_i\Vert)+\psi_i+\begin{bmatrix}
		w_i-\hat w_i\\
		h_{2i}(x_i)-\hat h_{2i}(\hat x_i)
		\end{bmatrix}^T\begin{bmatrix}
		\bar X_i^{11}&\bar X_i^{12}\\
		\bar X_i^{21}&\bar X_i^{22}
		\end{bmatrix}\begin{bmatrix}
		w_i-\hat w_i\\
		h_{2i}(x_i)-\hat h_{2i}(\hat x_i)
		\end{bmatrix}\bigg)
		\\\notag
		&=\sum_{i=1}^N\mu_i\kappa_iS_i( x_i,\hat x_i)+\sum_{i=1}^N\mu_i\rho_{\mathrm{ext}i}(\Vert \hat \nu_i\Vert)+\sum_{i=1}^N\mu_i\psi_i
		\\\notag
		&\quad+\begin{bmatrix}
		w_1-\hat w_1\\
		\vdots\\
		w_N-\hat w_N\\
		h_{21}(x_1)-\hat h_{21}(\hat x_1)\\
		\vdots\\
		h_{2N}(x_N)-\hat h_{2N}(\hat x_N)
		\end{bmatrix}^T\begin{bmatrix}
		\mu_1\bar X_1^{11}&&&\mu_1\bar X_1^{12}&&\\
		&\ddots&&&\ddots&\\
		&&\mu_N\bar X_N^{11}&&&\mu_N\bar X_N^{12}\\
		\mu_1\bar X_1^{21}&&&\mu_1\bar X_1^{22}&&\\
		&\ddots&&&\ddots&\\
		&&\mu_N\bar X_N^{21}&&&\mu_N\bar X_N^{22}
		\end{bmatrix}\begin{bmatrix}
		w_1-\hat w_1\\
		\vdots\\
		w_N-\hat w_N\\
		h_{21}(x_1)-\hat h_{21}(\hat x_1)\\
		\vdots\\
		h_{2N}(x_N)-\hat h_{2N}(\hat x_N)
		\end{bmatrix}\\\notag
		&=\sum_{i=1}^N\mu_i\kappa_iS_i( x_i,\hat x_i)+\sum_{i=1}^N\mu_i\rho_{\mathrm{ext}i}(\Vert \hat \nu_i\Vert)+\sum_{i=1}^N\mu_i\psi_i\\\notag
		&\quad+\begin{bmatrix}
		M\begin{bmatrix}
		h_{21}(x_1)\\
		\vdots\\
		h_{2N}(x_N)
		\end{bmatrix}- \hat M\begin{bmatrix}
		\hat h_{21}(\hat x_1)\\
		\vdots\\
		\hat h_{2N}(\hat x_N)
		\end{bmatrix}\\
		h_{21}(x_1)-\hat h_{21}(\hat x_1)\\
		\vdots\\
		h_{2N}(x_N)-\hat h_{2N}(\hat x_N)
		\end{bmatrix}^T\!\bar X_{cmp}\begin{bmatrix}
		M\begin{bmatrix}
		h_{21}(x_1)\\
		\vdots\\
		h_{2N}(x_N)
		\end{bmatrix}- \hat M\begin{bmatrix}
		\hat h_{21}(\hat x_1)\\
		\vdots\\
		\hat h_{2N}(\hat x_N)
		\end{bmatrix}\\
		h_{21}(x_1)-\hat h_{21}(\hat x_1)\\
		\vdots\\
		h_{2N}(x_N)-\hat h_{2N}(\hat x_N)
		\end{bmatrix}\\\notag
		&=\sum_{i=1}^N\mu_i\kappa_iS_i( x_i,\hat x_i)+\sum_{i=1}^N\mu_i\rho_{\mathrm{ext}i}(\Vert \hat \nu_i\Vert)+\sum_{i=1}^N\mu_i\psi_i\\\notag
		&\quad+\begin{bmatrix}
		h_{21}(x_1)-\hat h_{21}(\hat x_1)\\
		\vdots\\
		h_{2N}(x_N)-\hat h_{2N}(\hat x_N)
		\end{bmatrix}^T\begin{bmatrix}
		M\\
		\mathds{I}_{\tilde q}
		\end{bmatrix}^T\!\bar X_{cmp}\begin{bmatrix}
		M\\
		\mathds{I}_{\tilde q}
		\end{bmatrix}\begin{bmatrix}
		h_{21}(x_1)-\hat h_{21}(\hat x_1)\\
		\vdots\\
		h_{2N}(x_N)-\hat h_{2N}(\hat x_N)
		\end{bmatrix}\\\label{Eq_4a}
		&\leq\sum_{i=1}^N\mu_i\kappa_iS_i( x_i,\hat x_i)+\sum_{i=1}^N\mu_i\rho_{\mathrm{ext}i}(\Vert \hat \nu_i\Vert)+\sum_{i=1}^N\mu_i\psi_i\leq\kappa V\left( x,\hat{x}\right)+\rho_{\mathrm{ext}}(\left\Vert \hat \nu\right\Vert)+\psi.
		\end{align}
		\rule{\textwidth}{0.1pt}
	\end{figure*} 
\end{proof}

\begin{proof}\textbf{(Theorem~\ref{Thm_3a})}
Since $\hat C_1 =C_1P$, we have $\Vert C_1x-\hat C_1\hat x\Vert^2=(x-P\hat x)^TC_1^TC_1(x-P\hat x)$. Since  $\lambda_{\min}(C_1^TC_1)\Vert x- P\hat x\Vert^2\leq(x- P\hat x)^TC_1^TC_1(x- P\hat x)\leq\lambda_{\max}(C_1^TC_1)\Vert x- P\hat x\Vert^2$, and similarly  $\lambda_{\min}(\mathcal{\bar M})\Vert x- P\hat x\Vert^2\leq(x-P\hat x)^T \mathcal{\bar M}(x-P\hat x)\leq\lambda_{\max}(\mathcal{\bar M})\Vert x- P\hat x\Vert^2$, it can be readily verified that  $\frac{\lambda_{\min}(\mathcal{\bar M})}{\lambda_{\max}(C_1^TC_1)}\Vert C_1x-\hat C_1\hat x\Vert^2\le S(x,\hat x)$ holds $\forall x\in X$, $\forall \hat x\in \hat X$, implying that the inequality~\eqref{eq:V_dec1} holds with $\alpha(s)=\frac{\lambda_{\min}(\mathcal{\bar M})}{\lambda_{\max}(C_1^TC_1)}s^2$, $\forall s\in\R_{\geq0}$. 

We proceed with showing that the inequality~\eqref{eq:V_dec} holds, as well. Using Assumption~\ref{ASM: 1}, we have

\begin{align}\notag
\EE&\Big[S(\xi((k+1)\tau),\hat{\xi}(k+1))\,\big|\,\xi=\xi(k\tau),\hat \xi=\hat \xi(k),\nu = \nu(k\tau),\hat \nu= \hat \nu(k),w = w(k\tau),\hat w= \hat w (k)\Big]\\\notag
&=\EE \Big[S(\xi((k+1)\tau),\hat{\xi}(k+1))\,\big|\,\xi,\hat \xi,\nu,\hat \nu, w, \hat w\Big]-\EE \Big[S(\xi((k+1)\tau),\hat{\xi})\big|\xi,\hat \xi,\nu,\hat \nu, w, \hat w\Big]\\\notag
&\quad+\EE \Big[S(\xi((k+1)\tau),\hat{\xi})\,\big|\,\xi,\hat{\xi},\nu,\hat \nu, w, \hat w\Big]\\\notag
&\leq\EE \Big[S(\xi((k+1)\tau),\hat{\xi})\,\big|\,\xi,\hat{\xi},\nu,\hat\nu,w, \hat w\Big]+\EE \Big[\gamma(\Vert \hat \xi(k+1) - \hat \xi\Vert)\,\big|\,\hat\xi,\hat \nu, \hat w\Big]. 
\end{align}

Now by employing Dynkin's formula~\cite{dynkin1965markov}, one obtains

\begin{align}\notag
\EE& \Big[S(\xi((k+1)\tau),\hat{\xi})\,\big|\,\xi,\hat{\xi}, \nu,\hat\nu, w, \hat w\Big]+\EE \Big[\gamma(\Vert \hat \xi(k+1) - \hat \xi\Vert)\,\big|\,\hat{\xi},\hat \nu, \hat w\Big]\\\notag
&=\EE_{\varsigma} \Big[S(\xi,\hat \xi)+\EE\Big[\int_{k\tau}^{(k+1)\tau}\!\mathcal{L}S(\xi(t),\hat \xi)\mathsf{d}t \Big]\big|\hat{\xi},\hat \nu, \hat w\Big]+\EE \Big[\gamma(\Vert \hat \xi(k+1) - \hat \xi\Vert)\big|\hat{\xi},\hat \nu, \hat w\Big].
\end{align}

Since the \emph{infinitesimal generator} $\mathcal{L}S$, acting on the function $S$, is defined as
\begin{align}\label{infinitesimal generator}
\mathcal{L}S(\xi,\hat \xi)=\partial_\xi S(\xi,\hat \xi)f(\xi,\nu, w) + \frac{1}{2}\text{Tr}(\sigma(\xi)\sigma(\xi)^T\partial_{\xi,\xi}S(\xi,\hat \xi)),
\end{align}
where
\begin{align}\notag
\partial_\xi S(\xi,\hat \xi) = 2(\xi(t)-P\hat \xi)^T\mathcal{\bar M},\quad \partial_{\xi,\xi}S(\xi,\hat \xi) = 2\mathcal{\bar M},
\end{align}
one has
\begin{align}\notag
\EE_{\varsigma}	&\Big[S(\xi,\hat \xi)+\EE\Big[\int_{k\tau}^{(k+1)\tau}\!\mathcal{L}S(\xi(t),\hat \xi)\mathsf{d}t \Big]\,\big|\,\hat{\xi},\hat \nu, \hat w\Big]+\EE \Big[\gamma(\Vert \hat \xi(k+1)-\hat \xi\Vert)\big|\hat{\xi},\hat \nu, \hat w\Big]\\\notag
&=\EE_{\varsigma} \Big[S(\xi,\hat \xi)+\EE\Big[\int_{k\tau}^{(k+1)\tau}\!(2(\xi(t)-P\hat \xi)^T\mathcal{\bar M}(A\xi(t)+B\nu(t)+D w(t)+\mathbf{b})\\\notag
&\quad+G^T\mathcal{\bar M}G)\mathsf{d}t\Big]\,\big|\,\hat{\xi},\hat \nu, \hat w\Big]+ \EE \Big[\gamma\,(\Vert \,\hat \xi(k+1)-\hat \xi\,\Vert)\,\big| \hat{\xi},\hat \nu, \hat w\Big].
\end{align}   
Given any $\xi(t)$, $\hat \xi(k)$, $w(t)$ and $\hat w(k)$, we choose $\nu(t)$ via the following \emph{interface} function: 
\begin{align}\label{Eq_405a}
\nu(t) &= K(\xi(t)-P\hat \xi(k))-Q\hat \xi(k)
+(\xi(k\tau)-P\hat \xi(k))+H (w (k \tau)-\hat w(k) )- H w (t),
\end{align}
where $k\tau\leq t\leq(k +1)\tau$. By employing conditions~\eqref{Con_2}, and~\eqref{Con_3}, and the definition of the \emph{interface} function in~\eqref{Eq_405a}, we have
\begin{align}\notag
\EE_{\varsigma}& \Big[S(\xi,\hat \xi)+\EE\Big[\int_{k\tau}^{(k+1)\tau}\!(2(\xi(t)-P\hat \xi)^T\mathcal{\bar M}(A\xi(t)+B\nu(t)+ D w(t)+\mathbf{b})+G^T\mathcal{\bar M}G)\mathsf{d}t\Big]\,\big|\,\hat{\xi},\hat \nu, \hat w\Big]\\\notag
&\quad+\EE \Big[\gamma(\Vert \hat \xi(k+1)-\hat \xi\Vert)\,\big|\hat{\xi},\hat \nu, \hat w\Big]\\\notag
&=\EE_{\varsigma} \Big[S(\xi,\hat \xi)+\EE\Big[\int_{k\tau}^{(k+1)\tau}\!(2\,(\xi(t)-P\hat \xi)^T\mathcal{\bar M}((A+BK)(\xi(t)-P\hat \xi)+B(\xi(k\tau)-P\hat \xi(k))+ D(w-\hat w)+\mathbf{b})\\\notag
&\quad+G^T\mathcal{\bar M}G)\mathsf{d}t\Big]\big|\,\hat{\xi},\hat \nu, \hat w\Big]+\EE \Big[\gamma(\Vert \hat \xi(k+1)-\hat \xi\Vert)\big|\hat{\xi},\hat \nu, \hat w\Big].
\end{align}
Using Young's inequality~\cite{young1912classes} as $ab\leq \frac{\pi}{2}a^2+\frac{1}{2\pi}b^2,$ for any $a,b\geq0$ and any $\pi>0$, by employing Cauchy-Schwarz inequality and using the condition~\eqref{Con_1}, one has
\begin{align}\notag
\EE_{\varsigma}& \Big[S(\xi,\hat \xi)+\EE\Big[\int_{k\tau}^{(k+1)\tau}\!(2\,(\xi(t)-P\hat \xi)^T\mathcal{\bar M}((A+BK)(\xi(t)-P\hat \xi)+B(\xi(k\tau)-P\hat \xi(k))+ D(w-\hat w)+\mathbf{b})\\\notag
&\quad+G^T\mathcal{\bar M}G)\mathsf{d}t\Big]\big|\,\hat{\xi},\hat \nu, \hat w\Big]+\EE \Big[\gamma(\Vert \hat \xi(k+1) - \hat \xi\Vert)\big|\hat{\xi},\hat \nu, \hat w\Big]\\\notag
&\leq\EE_{\varsigma} \Big[S(\xi,\hat \xi)+\EE\Big[\int_{k\tau}^{(k+1)\tau}\!(-\tilde \kappa S(\xi(t),\hat \xi)+\pi\Vert \sqrt{\mathcal{\bar M}}\mathbf{b}\Vert^2+\pi \Vert \sqrt{\mathcal{\bar M}}B(\xi(k\tau)-P\hat \xi(k))\Vert^2+\pi \Vert \sqrt{\mathcal{\bar M}}D(w-\hat w)\Vert^2\\\notag
&\quad+ G^T\mathcal{\bar M}G)\mathsf{d}t\Big]\big|\,\hat{\xi},\hat \nu, \hat w\Big]+\EE \Big[\gamma(\Vert \hat \xi(k+1)-\hat \xi\Vert)\,\big|\,\hat{\xi},\hat \nu, \hat w\Big]\\\notag
&=\EE_{\varsigma} \Big[S(\xi,\hat \xi)+\EE\Big[\int_{k\tau}^{(k+1)\tau}\!\!\!-\tilde \kappa S(\xi(t),\hat \xi)\,\mathsf{d}t+\tau(\pi \Vert \sqrt{\mathcal{\bar M}}\mathbf{b}\Vert^2+ \pi \Vert \sqrt{\mathcal{\bar M}}B(\xi(k\tau)\!-\!P\hat \xi(k))\Vert^2+\pi \Vert \sqrt{\mathcal{\bar M}}D(w-\hat w)\Vert^2\\\notag
&\quad+ G^T\mathcal{\bar M}G)\Big]\big|\,\hat{\xi},\hat \nu, \hat w\Big]+\EE\Big[\gamma(\Vert \hat \xi(k+1) - \hat \xi\Vert)\big|\hat\xi,\hat \nu, \hat w\Big].
\end{align}	
Using Gronwall inequality~\cite{gronwall1919note}, one has
\begin{align}\notag
\EE_{\varsigma}& \Big[S(\xi,\hat \xi)+\EE\Big[\int_{k\tau}^{(k+1)\tau}\!-\tilde \kappa S(\xi(t),\hat \xi)\mathsf{d}t+\tau(\pi \Vert \sqrt{\mathcal{\bar M}}\mathbf{b}\Vert^2+ \pi \Vert \sqrt{\mathcal{\bar M}}B(\xi(k\tau)-P\hat \xi(k))\Vert^2+\pi \Vert \sqrt{\mathcal{\bar M}}D(w-\hat w)\Vert^2\\\notag
&\quad+G^T\mathcal{\bar M}G)\Big]\big|\,\hat{\xi},\hat \nu, \hat w\Big]+\EE \Big[\gamma(\Vert \hat \xi(k+1) - \hat \xi\Vert)\big|\hat\xi,\hat \nu, \hat w\Big]\\\notag
&\leq\EE_{\varsigma} \Big[e^{-\tilde \kappa \tau}S(\xi,\hat \xi)+\EE\Big[e^{-\tilde \kappa \tau}\tau(\pi\Vert \sqrt{\tilde  M}\mathbf{b}\Vert^2+ \pi \Vert \sqrt{\mathcal{\bar M}}B(\xi(k\tau)-P\hat \xi(k))\Vert^2+\pi \Vert \sqrt{\mathcal{\bar M}}D(w-\hat w)\Vert^2\\\notag
&\quad+G^T\mathcal{\bar M}G)\Big]\big|\,\hat{\xi},\hat \nu, \hat w\Big]+\EE \Big[\gamma(\Vert \hat \xi(k+1) - \hat \xi\Vert)\,\big|\hat\xi,\hat \nu, \hat w\Big]\\\notag
&=e^{-\tilde \kappa \tau}S(\xi,\hat \xi)+e^{-\tilde \kappa \tau}\tau (G^T\mathcal{\bar M}G+\pi\Vert \sqrt{\mathcal{\bar M}}\mathbf{b}\Vert^2+\pi \Vert \sqrt{\mathcal{\bar M}}B(\xi(k\tau)-P\hat \xi(k))\Vert^2+\pi \Vert \sqrt{\mathcal{\bar M}}D(w-\hat w)\Vert^2)\\\notag
&\quad+\EE \Big[\gamma(\Vert \hat \xi(k+1) - \hat \xi\Vert)\big|\hat\xi,\hat \nu, \hat w\Big]\\\notag
&=e^{-\tilde \kappa \tau}S(\xi,\hat \xi)+e^{-\tilde \kappa \tau}\tau (G^T\mathcal{\bar M}G+\pi\Vert \sqrt{\mathcal{\bar M}}\mathbf{b}\Vert^2)+\EE \Big[\gamma(\Vert \hat \xi(k+1) - \hat \xi\Vert)\big|\hat\xi,\hat \nu, \hat w\Big]\\\notag
&\quad+ \begin{bmatrix}\xi-P\hat \xi\\w-\hat w\\\end{bmatrix}^T \begin{bmatrix}
\pi e^{-\tilde \kappa \tau}\tau B^T\mathcal{\bar M}B&&0 \\
0&& \pi e^{-\tilde \kappa \tau}\tau D^T \mathcal{\bar M}D\\
\end{bmatrix}\begin{bmatrix}\xi-P\hat \xi\\w-\hat w\\\end{bmatrix}\!.
\end{align}
Employing~\eqref{Eq_8a} and since $\hat C_2 = C_2P$, we have
\begin{align}\notag
&e^{-\tilde \kappa \tau}S(\xi,\hat \xi)+e^{-\tilde \kappa \tau}\tau (G^T\mathcal{\bar M}G+\pi\Vert \sqrt{\mathcal{\bar M}}\mathbf{b}\Vert^2)+\EE \Big[\gamma(\Vert \hat \xi(k+1) - \hat \xi\Vert)\big|\hat\xi,\hat \nu, \hat w\Big]\\\notag
&\quad+ \begin{bmatrix}\xi-P\hat \xi\\w-\hat w\\\end{bmatrix}^T \begin{bmatrix}
\pi e^{-\tilde \kappa \tau}\tau B^T\mathcal{\bar M}B&&0 \\
0&& \pi e^{-\tilde \kappa \tau}\tau D^T \mathcal{\bar M}D\\
\end{bmatrix}\begin{bmatrix}\xi-P\hat \xi\\w-\hat w\\\end{bmatrix}\\\notag
&\leq e^{-\tilde \kappa \tau}S(\xi,\hat \xi)+e^{-\tilde \kappa \tau}\tau (G^T\mathcal{\bar M}G+\pi\Vert \sqrt{\mathcal{\bar M}}\mathbf{b}\Vert^2)+\EE \Big[\gamma(\Vert \hat \xi(k+1) - \hat \xi\Vert)\big|\hat\xi,\hat \nu, \hat w\Big]\\\notag
&\quad+ \begin{bmatrix}\xi-P\hat \xi\\w-\hat w\\\end{bmatrix}^T \begin{bmatrix}
\bar\kappa \mathcal{\bar M}+C_2^T\bar X^{22}C_2 & C_2^T\bar X^{21}\\
\bar X^{12}C_2 & \bar X^{11}\\
\end{bmatrix}\begin{bmatrix}\xi-P\hat \xi\\w-\hat w\\\end{bmatrix}\\\notag
& = (\bar\kappa+e^{-\tilde \kappa \tau})S(\xi,\hat \xi)+e^{-\tilde \kappa \tau}\tau (G^T\mathcal{\bar M}G+\pi\Vert \sqrt{\mathcal{\bar M}}\mathbf{b}\Vert^2)+\EE \Big[\gamma(\Vert \hat \xi(k+1) - \hat \xi\Vert)\big|\hat\xi,\hat \nu, \hat w\Big]\\\notag
&\quad+ \begin{bmatrix}w-\hat w\\C_2\xi-\hat C_2\hat \xi\\\end{bmatrix}^T \begin{bmatrix}
\bar X^{11} & \bar X^{12}\\
\bar X^{21} & \bar X^{22}\\
\end{bmatrix}\begin{bmatrix}w-\hat w\\C_2\xi-\hat C_2\hat \xi\\\end{bmatrix}\!.
\end{align}

Since the function $\gamma$ defined in Assumption~\ref{ASM: 1} is \emph{concave}, using Jensen inequality one has
\begin{align}\notag
\EE&\Big[\gamma(\Vert \hat \xi(k+1) - \hat\xi\Vert)\,\big|\,\hat\xi,\hat \nu, \hat w\Big]=\EE\Big[\gamma(\Vert \hat \xi(k+1) - (\hat \xi+ \hat \nu + \tilde D \, \hat w+\tilde R\varsigma)+ (\hat \xi+ \hat \nu + \tilde D \, \hat w+\tilde R\varsigma)- \hat \xi\Vert)\,\big|\,\hat\xi,\hat \nu, \hat w\Big]\\\notag
&\leq \EE\Big[\gamma(\delta+\Vert \hat \nu+ \tilde D\hat w+\tilde R\varsigma\Vert)\,\big|\,\hat\xi,\hat \nu, \hat w\Big]\\\notag
&\le\gamma((1+\bar\eta)\delta)+\EE\Big[\gamma((1+\frac{1}{\bar\eta})\Vert \hat \nu+ \tilde D \hat w+\tilde R\varsigma\Vert)\,\big|\,\hat\xi,\hat \nu, \hat w\Big]\\\notag
&\le\gamma((1+\bar\eta)\delta)+\gamma((1+\frac{1}{\bar\eta})(1+\bar\eta')\Vert \hat \nu+ \tilde D\, \hat w\Vert)+\gamma((1+\frac{1}{\bar\eta})(1+\frac{1}{\bar\eta'})\EE\Big[\Vert \tilde R\varsigma\Vert\,\big|\,\hat\xi,\hat \nu, \hat w\Big])\\\notag
&=\gamma((1+\bar\eta)\delta)+\gamma((1+\frac{1}{\bar\eta})(1+\bar\eta')\Vert \hat \nu + \tilde D \, \hat w\Vert)+\gamma((1+\frac{1}{\bar\eta})(1+\frac{1}{\bar\eta'})\EE\Big[([\tilde R\varsigma]^T[\tilde R\varsigma])^{\tfrac{1}{2}}\big|\,\hat\xi,\hat \nu, \hat w\Big])\\\notag
&\le\gamma((1+\bar\eta)\delta)+\gamma((1+\frac{1}{\bar\eta})(1+\bar\eta')(1+\bar\eta'')\Vert \hat \nu\Vert)+\gamma((1+\frac{1}{\bar\eta})(1+\bar\eta')(1+\frac{1}{\bar\eta ''})\Vert \tilde D \,\Vert\, \Vert \hat w\Vert)\\\notag
&\quad+\gamma((1+\frac{1}{\bar\eta})(1+\frac{1}{\bar\eta'})(\EE\Big[[\tilde R\varsigma]^T[\tilde R\varsigma]\,\big|\,\hat\xi,\hat \nu, \hat w\Big])^{\tfrac{1}{2}})\\\notag
&=\gamma((1+\bar\eta)\delta)+\gamma((1+\frac{1}{\bar\eta})(1+\bar\eta')(1+\bar\eta'')\Vert \hat \nu\Vert)+\gamma((1+\frac{1}{\bar\eta})(1+\bar\eta')(1+\frac{1}{\bar\eta ''})\,\Vert \tilde D  \,\Vert\, \Vert \hat w\Vert)\\\label{norm_gamma}
&\quad+\gamma((1+\frac{1}{\bar\eta})(1+\frac{1}{\bar\eta'})\sqrt{\text{Tr}(\tilde R^T\tilde R)}),
\end{align}
where $\bar\eta,\bar\eta', \bar\eta'' \in\mathbb R_{>0}$. 	
Then one can conclude that	

\begin{align}\notag
\EE& \Big[S(\xi((k+1)\tau),\hat{\xi}(k+1))\,\big|\,\xi,\hat{\xi},\nu,\hat \nu, w, \hat w\Big]\\\notag
&\leq (\bar\kappa+e^{-\tilde \kappa \tau})S(\xi,\hat \xi)+\gamma((1+\frac{1}{\bar\eta})(1+\bar\eta')(1+\bar\eta'')\Vert \hat \nu\Vert)+e^{-\tilde \kappa \tau}\tau (G^T\mathcal{\bar M}G+\pi\Vert \sqrt{\mathcal{\bar M}}\mathbf{b}\Vert^2)+\gamma((1+\bar\eta)\delta)\\\notag
&\quad+\gamma((1+\frac{1}{\bar\eta})(1+\frac{1}{\bar\eta'})\sqrt{\text{Tr}(\tilde R^T\tilde R)})+\gamma((1+\frac{1}{\bar\eta})(1+\bar\eta')(1+\frac{1}{\bar\eta ''}) \,\Vert \tilde D \,\Vert\, \Vert \hat w\Vert)\\\label{error}
&\quad+ \begin{bmatrix}w-\hat w\\C_2\xi-\hat C_2\hat \xi\\\end{bmatrix}^T \begin{bmatrix}
\bar X^{11} & \bar X^{12}\\
\bar X^{21} & \bar X^{22}\\
\end{bmatrix}\begin{bmatrix}w-\hat w\\C_2\xi-\hat C_2\hat \xi\\\end{bmatrix}\!,
\end{align}
which completes the proof with
\begin{align}\notag
\alpha(s)&:=\frac{\lambda_{\min}(\mathcal{\bar M})}{\lambda_{\max}(C_1^TC_1)}s^2, \quad \forall s\in\R_{\geq0},\\\notag
\kappa &:= \bar\kappa+e^{-\tilde \kappa \tau},\\\notag
\rho_{\mathrm{ext}}(s) &:= \gamma((1+\frac{1}{\bar\eta})(1+\bar\eta')(1+\bar\eta'') s), \quad \forall s\in\R_{\geq0},\\\notag
\psi &:=e^{-\bar \kappa \tau}\tau (G^T\mathcal{\bar M}G+\pi\Vert \sqrt{\mathcal{\bar M}}\mathbf{b}\Vert^2)+\gamma((1+\bar\eta)\delta)+\gamma((1+\frac{1}{\bar\eta})(1+\frac{1}{\bar\eta'})\sqrt{\text{Tr}(\tilde R^T \tilde R)})\\\notag
&\quad+\gamma((1+\frac{1}{\bar\eta})(1+\bar\eta')(1+\frac{1}{\bar\eta ''}) \,\Vert \tilde D \,\Vert\, \Vert \hat w\Vert).
\end{align}
\end{proof}

\begin{remark}
	Note that the infinitesimal generator $\mathcal{L}S(x,\hat x)$ defined in~\eqref{infinitesimal generator} is different from the usual one employed in~\cite{zamani2014symbolic} since the abstract system $\widehat\Sigma$ in this paper is considered in the \emph{discrete-time} domain. 
\end{remark}

\end{document}